# Minecraft: An Engaging Platform to Learn Programming

Worasait Suwannik
Department of Computer Science
Kasetsart University
Bangkok, Thailand
worasait.suwannik@gmail.com

*Abstract*—**Teaching programming effectively is difficult. This paper explores the benefits of using Minecraft Education Edition to teach Python programming. Educators can use the game to teach various programming concepts ranging from fundamental programming concepts, object-oriented programming, event-driven programming, and parallel programming. It has several benefits, including being highly engaging, sharpen creativity and problem-solving skill, motivating the study of mathematics, and making students realizes the importance of programming.**

*Keywords-Minecraft; teaching programming*

I. INTRODUCTION

Coding is an important skill not only for a knowledge worker in IT or CS industries but also in other industries. Non-CS workers who are able to code have an advantage over those who cannot. An engineer can solve a more complex problem by writing custom calculation code for a task that is specific, new, or extended from an existing task. An accountant can gain benefits from analyzing financial data. An artist can be more productive by generating thousands of images.

However, programming is difficult. Programming syntax is rigorous. For example, a Python interpreter will give an error for not indenting property, missing an enclosing parenthesis, or having an extra semicolon. Translating word problems to code is even more difficult. It deals with abstract and logical thinking. Understanding error messages requires experience or help from the community. Debugging is even harder than writing a program.

There are many approaches to overcome this problem. The concept of coding can be learned using games such as Lightbot. The game simplifies several aspects of programming. For example, the syntax is reduced to a few robot movement commands; no conditional statement; looping is replaced by a simple procedure call. The only task the program needs to do is to light specific tiles on different maps.

Another approach is to motivate students to program. For some people, programming is very engaging because a computer can give instant feedback on logical errors or misunderstandings of problems. However, some people do not engage with data processing on their computers. One study suggests using tangible wearable and robotics platforms to engage learners [1].

This paper presents Minecraft Education Edition as another tool that can engage students in learning programming. Minecraft Education Edition is based on Minecraft, which is the number 1 ranking video game [2]. Even though Minecraft is not a tangible platform, students can interact with structures built from code.

II. MINECRAFT EDUCATION EDITION

Minecraft is a 3D video game developed by Mojang Studios, which Microsoft owns since 2014. There are many ways to play Minecraft. A player can build structures using various blocks much like how we play Lego. Since Minecraft is a multiplayer game, a player can invite others to visit the structure or collaboratively build a structure using their server or they can subscribe and join business-hosted Minecraft servers.

Minecraft for Education is a game-based learning platform. A player signs in using an institution Microsoft account. For camps, clubs, and other organizations, licenses can be purchased if they have a Microsoft 365 Admin Center account [3]. Minecraft for Education is available for Windows, Mac, iPad, and Chromebook. This game helps educators in engaging students to learn various subjects: including chemistry, art, history, climate, social, and coding.

III. PROGRAMMING CONCEPTS

Minecraft for Education support Blocks, Python, JavaScript. However, the iPad version supports only Block and Python. A player can write code to build a structure or program an agent to do something for them, such as closing a door (see item 1 in Table 1). Table 1 lists examples of using Minecraft Education Edition to build structures in the game. The examples are posted in our YouTube Craft of Code channel and the code is stored in our github repository https://github.com/base0/minecraft To reference an example, we use # followed by the number of the example in Table 1. For example, #3 refers to the example number 3 in Table 1.

There are 3 editors in Minecraft Education Edition: MakeCode, Python notebook interface, and Tynker. MakeCode supports visual programming (Fig 1) and traditional text-based programming (Fig 2). For text-based programming, players can drag a code snippet from the left panel to the text area. In addition, it has a code completion feature, which pops up a list of methods, constants, variables so that a player does not have to type or to memorize the exact name. This paper focuses on MakeCode's traditional text-based Python programming.

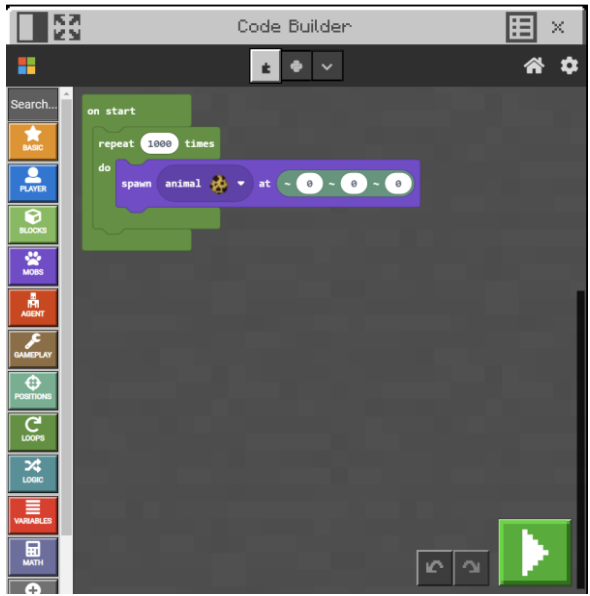

Figure 1.  MakeCode Block editor

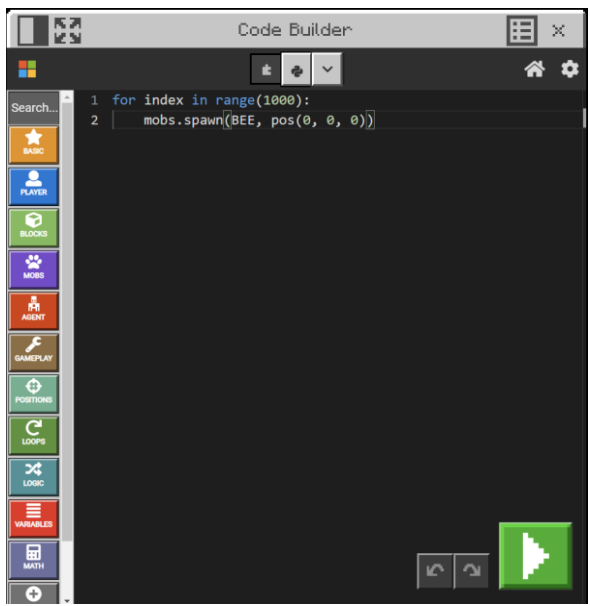

Figure 2.  MakeCode Python editor

### A. Fundamental Programming Concepts

Python in Minecraft MakeCode has fundamental programming constructs such as a sequence, a condition, a loop, and a function.

In the current version of Minecraft Education Edition (17.8), Python in Minecraft MakeCode is different from desktop Python. It does not support Python list comprehension. The error message is "Cannot find name 'ListComp'". This version lacks some built in functions such as all, dir, type. However, the missing syntax and functions are available in Notebook interface.

### B. Object-Oriented Programming

Object-oriented programming (OOP) is a popular programming language paradigm. According to August 2022 TIOBE index [4], 8 of the top ten programming language supports OOP. In object-oriented programming, a programmer can combine related data and code into a class. This can make the program easy to maintain and debug. After a class is defined, objects of the class can be created. A class can be reused using an inheritance mechanism. Minecraft Education Edition supports object-oriented programming in Python.

### C. Event-Driven Programming

An event may come from user interaction with a GUI component, a hardware sensor, or incoming network traffic. In event-driven programming, the execution of code depends on event that the program receives. Event-driven programming is used ubiquitously in application development. It is used in mobile applications, desktop applications, and server applications.

A player can write an event-driven program in Minecraft for Education. A player can write code that responses to chat, character movement, mobs (i.e., a computer-controlled character in a game) died, using an item, etc.

### D. Parallel Programming.

In parallel programming, more than one part of a computer program can be executed simultaneously. This type of program can effectively make use of multicore CPUs or GPUs. In Minecraft, a parallel program can speed up the construction even more.

## IV. WHY MINECRAFT

This section describes benefits of using Minecraft for Education.

### A. Engaging

Minecraft Education Edition can be used to teach various programming concepts in a fun and engaging way.

- Loop.  A player can use a loop to spawn hundreds or thousands of animals (#2) (see Figure 1 and 2)

- Function. A function can be used to build one floor (#3), a house (#1), or an igloo (#4). The function can receive parameters to vary the construction location.

- Recursive. To explain a recursion concept, a Russian doll house is normally used as an analogy. We build a Russian doll house in Minecraft using a recursive call (#5)

- Event-Driven Programming. Event-driven programming is used to show A-Z letters when our character is moving (#6), build a lava wall that blocks zombies (#7), or building a staircase when walking forward (#8).

- Object-Oriented Programming. A player can inherit an existing code to make a structure more colorful (#21).

- Parallel Programming. An example of constructing with parallel programming is to plant one million roses in 3 minutes (#9).

After a program finish building a structure, a player can explore (#3), see how it looks at nighttime (#10), look at fish through the window of underwater house (#11), and have an experience of being inside a giant bottle (#12). Animals can be put in the structure (#13). We might observe how various animal interact with our structures. For example, how a bat can solve a maze (#14) or how cats climb up or down a giant pyramid (#15).

Minecraft Education Edition provides highly engaging programming exercise through story. For example, in Hour of Code 2021 (TimeCraft), a player goes back in time and use coding to save history from changing. For example, a player can help Ada fixing her punch card. In addition, while solving history mishaps, a player collects clues that can identifies the time culprit.

### B. Amplifying Creativity

Creativity is an essential skill in for 21st century. It becomes even more critical as more and more tasks can be performed by artificial intelligence. A player can sharpen their creativity skill using Minecraft. In Minecraft, a player can build structure on the ground, in the sky (#16), inside a mountain (#17), or underwater (#11) The possibility is unlimited. With coding, the possibility is increased even more. This is because a player can build more complicate structures with code or build things based on an event, such as imitating the scene from the Frozen movie when queen Elsa ran and used her magical power to create icy stair steps (#8)

### C. Sharpen a Problem Solving Skill

A student must turn the problem statement into code in a programming exam or a programming exercise. In Minecraft, a player has an idea of what they want to build. If they are going to build with code, they must translate the structure into code.

During the process, a player might have to do some research. For example, in building an aquarium, a Sandstone block is needed to make bubbles. To build a room with a lot of black cats, a player might need to come up with a construction that can quickly select only black cats from other cats (#20).

### D. Motivation to Study Mathematics

To build a particular structure, a specific mathematics might be required. For example, building a leaning tower requires mathematics to rotate a tower (#18). Curling a dragon around a pole requires trigonometry (#19) Therefore, a player is motivated to study mathematics in order to build those structures.

### E. Demonstrate the Importance of Coding

For a large structure, building it manually take a long time. Coding help player building those structure much quicker than doing it manually. As Gabe said "The programmers of tomorrow are the wizards of the future. You gonna look like you have magic powers compared to anybody else." In Minecraft, those who can code can do more and can manipulate the world such as change the time and climate, or teleport to another location much like a wizard. They can even write an event-driven program that responses to a chat event to block zombies using a lava wall (#7), similar to when a wizard is casting a spell.

## V. CONCLUSION

Minecraft can be used to teach various programming concepts such as recursion, event-driven programming, and parallel programming. It is very engaging because player can interact with the structure built from code. The programmer teaching module provides story. Players can sharpen their creativity and problem-solving skills. Building structure in Minecraft Education Edition with code motivate players to study mathematics and reinforce the importance of coding over doing thing manually.

## ACKNOWLEDGMENT

The author would like to thank Assoc. Prof. Somying Thainimit for explaining the uses of programming in engineering.

| No | Title | YouTube Video Id |
|---|---|---|
| 1 | 100-house village | hI911-kK4a4 |
| 2 | Birds | r8_Ufbalec4 |
| 3 | 20-storey building | 5g46_MTbch4 |
| 4 | 50-igloo village | SWOBWXOsCUY |
| 5 | Russian House using Recursion | RbaaTgJAnOI |
| 6 | Minecraft ABC | Rc29AuuVBKo |
| 7 | zombie attack | NaaN2haefPI |
| 8 | Creating stair steps while walking in the snow inspired by Frozen Let It Go | 5HwyshXx2tk |
| 9 | 1,000,000 roses in under 3 minutes | E5XFEYgONB4 |
| 10 | bee hive | P4xN5A2i5rY |
| 11 | underwater house | BYXTgNNIjBw |
| 12 | what was it like to be inside a soda bottle? | 3v6QDFXNR |
| 13 | Maze | FnwBV-8xNWM |
| 14 | can bats solves an air maze at nighttime? | mTtCzETSlr8 |
| 15 | Extreme pyramid | 0QdXh7hHHcA |
| 16 | celestial castle | m11OnDVT9CM |
| 17 | two-storey cave | K2epbunPjTY |
| 18 | LEANING tower | u2tIqfDbIyU |
| 19 | chinese new year dragon | 7DONth96xmI |
| 20 | black cats superstitious | oQ2tejU_D14 |
| 21 | colorful spheres | FlCRH5KW2Ak |